\documentclass[12pt]{article}

\usepackage[pdftex]{graphicx}
\usepackage{cite}

\usepackage{multirow}%
\usepackage{amsmath,amssymb,amsfonts}%
\usepackage{amsthm}%
\usepackage{mathrsfs}%
\usepackage{xcolor}%
\usepackage{authblk}
\usepackage{hyperref}
\usepackage{microtype}

\usepackage{hyperref}
\usepackage{lineno}
\usepackage{slashed} 
\usepackage{SpinorNotation/Spinor} 
\usepackage{SpinorNotation/DiSpinor} 
\usepackage{SpinorNotation/ReOperators}
\usepackage{SpinorNotation/Clifford}
\usepackage{SpinorNotation/StandardModel}
 
\newcommand{\sss}[1]{\scriptscriptstyle{#1}}
\newcommand{\ds }{\displaystyle}

\newcommand{\calB}{\mathcal{B}}
\newcommand{\calC}{\mathcal{C}}

\newcommand{\calG}{\mathcal{G}}
\newcommand{\spIU}[1]{u_{#1}}

\newcommand{\spVI}[1]{{\bar v_{#1}}}

\begin{document}

\title{EW one-loop corrections to the longitudinally polarized Drell-Yan process. \\ (II) Charged-current case.}

\author[1,5]{S.\,Bondarenko}
\author[2,3,5]{Ya.\,Dydyshka}
\author[2,4]{L.\,Kalinovskaya}
\author[2,4]{A.\,Kampf} 
\author[2]{R.\,Sadykov}
\author[2,3,5]{V.\,Yermolchyk}

\affil[1]{\small Bogoliubov Laboratory of Theoretical Physics, Joint Institute for Nuclear Research,
  Dubna 141980, Russia}
\affil[2]{\small Dzhelepov Laboratory of Nuclear Problems, Joint Institute for Nuclear Research, 
  Dubna, 141980, Russia}
\affil[3]{\small The Institute for Nuclear Problems, Belarusian State University, Minsk, 220030, Belarus}
\affil[4]{\small Lomonosov Moscow State University, Moscow, 119991, Russia}
\affil[5]{\small Dubna State University, Dubna, 141980, Russia}

\date{\today}

\maketitle

\abstract{
Complete one-loop electroweak radiative corrections to the charged-current Drell-Yan processes $pp \to \ell^{+}\nu_{\ell}(+X)$ and $pp \to \ell^{-}\bar{\nu}_{\ell}(+X)$
are presented for the case of longitudinal polarization of initial particles.
The results can be used to obtain precise predictions 
for the kinematic distributions of polarized $W^{\pm}$ production cross sections, and
single- and double-spin asymmetries.
Numerical results are obtained using the Monte-Carlo generator {\tt ReneSANCe}. This research contributes to a global next-to-leading order analysis of polarized parton distributions in proton-proton collisions at RHIC. 
We proved that the impact of the one-loop electroweak radiative corrections to single- and double-spin asymmetries is negligible.
}

\vspace*{6pt}

\label{sec:intro}
\section*{Introduction}

The Drell-Yan (DY) process \cite{Drell:1970wh} has become one of the most important processes for experimental investigations of $pp$-collisions due to the large production rates and clean experimental signatures.

The unpolarized case of the charged-current (CC) DY process was quite well studied at various levels of quantum chromodynamics (QCD) \cite{Balazs:1997xd,Ellis:1997sc,Ellis:1997ii,Anastasiou:2003ds,Melnikov:2006di,Melnikov:2006kv,Catani:2009sm}, one-loop electroweak  (EW)~\cite{Wackeroth:1996hzen,Baur:1998kt,Dittmaier:2001ay,CarloniCalame:2003ux,Baur:2004ig,CarloniCalame:2006zq,Arbuzov:2005dd,Brensing:2007qm,Bardin:2008fn} and mixed QCD-EW \cite{Dittmaier:2014qza,Dittmaier:2015rxo,Bonciani:2016wya,Behring:2020cqi,Buonocore:2021rxx,Armadillo:2024nwk} radiative corrections (RCs). Impact of one-loop EW RC for longitudinally polarized case was first investigated in \cite{Zykunov:2001mn,Zykunov:2003gm}. The single- and double-spin asymmetries are presented as a function of lepton transversal energy for several pseudo-rapidities.

To shed light on the spin structure of protons, it is necessary to study polarized 
case of CC DY process. It can be used 
for this purpose, since the $W^{\pm}$ bosons naturally choose left-hand quark and right-hand antiquark orientations. Thus, the expected parity violation will allow unique and precise measurements of the spin direction of quarks and antiquarks in the proton.
Knowledge about polarized parton distribution functions (PDFs) and the contribution of longitudinally polarized sea of quarks and gluons to the proton spin  can be extracted from single- and double-spin asymmetries. 

RHIC \cite{Bunce:2000uv,Aschenauer:2013woa,RHICSPIN:2023zxx} is a unique collider that can carry out experiments with the polarized high-energy beams of protons. To date, experiments have been performed at energies in the region of 500 GeV. The first experimental data on measuring asymmetries at $\sqrt{s} = 510$ GeV in the center-of-mass system (c.m.s.) were published recently \cite{RHICSPIN:2023zxx}. We see quite large uncertainties in experimental measurements (see Fig. 2 in \cite{RHICSPIN:2023zxx}), which are explained by rather small statistics. To provide theoretical support at this energy scale, a comprehensive analysis is needed, including evaluation of both QCD and EW RCs.

The motivation of the present research is to contribute to a global next-to-leading order (NLO) analysis of polarized parton distributions, including all available data from both lepton scattering and proton-proton collisions at RHIC. 
We calculated the CC DY process to determine longitudinally polarized cross sections and single- and double-spin asymmetries as functions of the lepton pseudo-rapidity and investigated influence of the one-loop EW RCs.
Previously in ~\cite{Bondarenko:2022zse}, we carried out similar investigation
for neutral current DY. 
One-loop QCD analysis for $pp$-scattering at RHIC was performed in~\cite{deFlorian:2008mr,deFlorian:2009vb,deFlorian:2010aa,deFlorian:2014yva} and the set of polarized PDF was created.
Numerical results are obtained using the Monte-Carlo (MC) generator {\tt ReneSANCe} \cite{Bondarenko:2022mbi}.

The paper is organized as follows. 
We briefly describe the cross section within the helicity approach at partonic and hadronic levels in Section \ref{section1}. In Section \ref{section2}  
we define the observables that are sensitive to polarization and are traditionally used to estimate it.
Section \ref{section3} is devoted to the presentation and discussion of
numerical results.
Our conclusions are drawn in Section \ref{Conclusion}.

\section{Differential cross section \label{section1}}

To study the case of longitudinal polarization, 
we calculate helicity amplitudes (HAs) and apply equation~(1.15) from~\cite{MoortgatPick:2005cw}.

\subsection{Hadronic level}
   
The production of a single $W^{\pm}$-bosons at the $pp$-colliders is given by the reaction:
\begin{eqnarray*}
& pp \to W^+ + X \to \ell^+\nu_\ell + X,\\ 
& pp \to W^- + X \to \ell^-\bar{\nu}_\ell + X,
\label{DYCC} 
\end{eqnarray*}
with $\ell= e,\mu$.

The differential cross section of
the DY process at the hadronic level can be obtained from the parton cross section by convolution with the corresponding PDFs:
\begin{eqnarray*}
\label{sigpp}
&& d\sigma(\Lambda_1,\Lambda_2,s) =
\sum_{q_1 q_2}\sum_{\lambda_1\lambda_2}\int_{0}^{1}\int_{0}^{1} dx_1 dx_2	
\\ && 
{f}_{q_1}^{\Lambda_1\lambda_1}(x_1) 
\times
{f}_{q_2}^{\Lambda_2\lambda_2}(x_2)\,
d\hat{\sigma}_{q_1 q_2}(\lambda_1,\lambda_2,\hat{s}) \nonumber
\end{eqnarray*}
where $\Lambda_i = \pm 1$ and $\lambda_i = \pm 1$ are the helicities of each proton and quark, respectively,
with $\hat{s} = x_1 x_2s$.
Parton distributions ${f}_{q_i}^{\Lambda_i\lambda_i}$ can be obtained from unpolarized
$f_{q_i}$ and longitudinally polarized $\Delta f_{q_i}$ PDFs
\begin{eqnarray*}
{f}_{q_i}^{\Lambda_i\lambda_i} = \frac{1}{2}( f_{q_i} + \Lambda_i\lambda_i\Delta f_{q_i}).
\end{eqnarray*}

\subsection{Partonic level}

At the partonic level we consider reactions in the following form: 
\begin{gather}
\begin{aligned}
\label{partonllL}
\bar{d}(p_1,\lambda_1)  +& 
  {u}(p_2,\lambda_2) \to
l^+(p_3,\lambda_3)  + 
\nu_l(p_4,\lambda_4)~(+ \gamma(p_5,\lambda_5)), \\  
\bar{u}(p_1,\lambda_1)  +& 
{d}(p_2,\lambda_2)\to  
{l}^-(p_3,\lambda_3) +  
\bar\nu_l(p_4,\lambda_4)~(+ \gamma(p_5,\lambda_5)).   
\end{aligned}
\end{gather}

\noindent 
Arguments indicate momenta $p_i$ and helicities $\lambda_i$ of initial and final particles. 

We first introduce the notation used in the presentation of our results at the one-loop level. As usually, we subdivide the differential cross section as follows:
\begin{eqnarray*}
\hat{\sigma}^{\text{1-loop}}  =  \hat{\sigma}^{\mathrm{Born}} + \hat{\sigma}^{\mathrm{virt}}(\lambda) + \hat{\sigma}^{\mathrm{soft}}(\lambda, \omega)  
+ \hat{\sigma}^{\mathrm{hard}}(\omega) + \hat{\sigma}^{\mathrm{Subt}}, 
\label{loopxsec}
\end{eqnarray*}
where $\hat{\sigma}^{\mathrm{Born}}$ is the contribution of the Born cross section,
$\hat{\sigma}^{\mathrm{virt}}$ is virtual (loop) corrections,
$\hat{\sigma}^{\mathrm{soft}}$ is the soft photon emission,
and $\hat{\sigma}^{\mathrm{hard}}$ is the hard photon emission part (with energy $E_{\gamma} > \omega$)
and is defined using the soft-hard separator $\omega$ along with the auxiliary parameter $\lambda$ (fictitious {\it photon mass} which regularizes infrared divergences).
The special term  $\hat{\sigma}^{\mathrm{Subt}}$ denotes the subtraction of collinear quark mass singularities.
When all the contributions to the  cross section are added, the result is free of infrared divergences.
The partonic cross section is taken in the c.m.s. of initial quarks/antiquarks.

We treat all contributions using the HAs approach
with sum over helicities of all final state particles.

\subsubsection{Subtraction of the quark mass singularities}

\noindent 
The subtraction procedure at the parton level is implemented in the same way as in~\cite{Arbuzov:2005dd}.
The one-loop RCs contain term $\hat{\sigma}^{\mathrm{Subt}}$, proportional to the logarithms of the quark masses $\ln(\hat{s}/m^2_{u,d})$. Such singularities are well known, and in the case of hadronic collisions they are already effectively accounted in the PDF functions. 

\subsubsection{Born and virtual contribution} 

\noindent 
The covariant amplitude (CA) can be represented as the sum of
independent invariant basis elements (structures) of the scattering matrix multiplied by invariant functions (form factors). The structures are kinematic objects and contain the entire spin dependence, whereas 
form factors are scalar quantities and carry information about the dynamics of the process. 

For the case when only lepton mass is not neglected
($m_l \neq 0$) we present the CA decomposed on a massive basis with 
the $LL$, $LLD$, and $LRD$ contributions.
The short expression for the CA  can be written as
\begin{eqnarray*}
    {\cal A}=
   i e^2 \dfrac{\chi_{\sss W}(\hat{s}) }{4\hat{s}} \sum_{i=1}^2
    Str_i {\cal F}_i(\hat{s},\hat{t},\hat{u}),
\end{eqnarray*}
where   
\begin{eqnarray*}
\chi_{\sss W}(\hat{s})=\dfrac{V_{12}}{2\SW^2}\dfrac{\hat{s}}{\hat{s}-\MW^2},
\end{eqnarray*}
$V_{12}$ is the relevant element of the CKM-matrix and $\SW$ is the sine of the Weinberg angle and $\hat s =(p_1+p_2)^2$, $\hat t =(p_1-p_3)^2$, $\hat u =(p_1-p_4)^2$ are Mandelstam variables.
The corresponding Dirac structures are  
\begin{eqnarray*}
Str_1 &=& Str_{\sss LL}=\gamma_\mu (1+\gamma_5)\otimes\gamma_\mu (1+\gamma_5), \\ \nonumber
Str_2 &=& Str_{\sss LLD}=\gamma_\mu (1+\gamma_5)\otimes\gamma_\mu (1+\gamma_5)(-i D_\mu)
\end{eqnarray*}
in the case of the $W^+$ channel
and
$$Str_2 = Str_{\sss LRD}=\gamma_\mu (1+\gamma_5)\otimes\gamma_\mu (1-\gamma_5)(-i D_\mu)$$
in the case of $W^{-}$ channel, where $D_\mu=(p_4-p_3)_\mu$.
Symbol $\otimes$ is used in the following brief notation:\\
 $\gamma_\mu \otimes \gamma^\nu =
 \bar v(p_1) \gamma_\mu u(p_2)
 \bar u(p_3) \gamma^\nu  v(p_4)$.
Scalar form factors $ {\cal F}_{\sss LL, LLD, LRD}$ are labeled according to their structures.
\noindent To obtain the corresponding HAs for the virtual part we apply an internal procedure {\tt SANC} based on the Vega-Wudka approach
\cite{Vega:1995cc}.
For both  channels two non-zero HAs survive. 
For the $W^+$ channel,  in the limit with of lepton mass $m_l^2=0$ they are
\begin{eqnarray*}
{\cal H}_{+--+}^{W^+} 
&=& -e^2 
(1+\cos\vartheta_{23})\chi_{\sss W}(\hat{s})\dfrac{1}{\sqrt{\hat{s}}} {\cal F}_{\sss LL},\\ \nonumber
{\cal H}_{+---}^{W^+} &=& -e^2 
\sin\vartheta_{23}
   \chi_{\sss W}(\hat{s}) 
   \left(\dfrac{m_l}{\hat{s}}{\cal F}_{\sss LL}+{\cal F}_{\sss LRD}\right),
\end{eqnarray*}
and for the $W^-$ channel, they are
\begin{eqnarray*}
{\cal H}_{+--+}^{W^-} &=&{\cal H}_{+--+}^{W^+}
, 
\nonumber \\
{\cal H}_{+-++}^{W^-} &=& 
{\cal H}_{+---}^{W^+}\left({\cal F}_{\sss LRD}\to {\cal F}_{\sss LLD}\right),
\end{eqnarray*}
where $\vartheta_{23}$ is the angle between particle 2 and 3 in reactions (\ref{partonllL}).

\subsubsection{Real photon emission}

Real photon emission consists of two parts: soft and hard photon Bremsstrahlung.

$\bullet$ The soft photon contribution $\hat{\sigma}^{\mathrm{soft}}(\lambda, \omega)$

The soft photon Bremsstrahlung contribution $\hat{\sigma}^{\mathrm{soft}}(\lambda, \omega)$ is factorized in front of the Born level cross section 
\begin{eqnarray*}
    \text{d}\hat{\sigma}^{\text{soft}}_{\lambda_{1}\lambda_{2}}(\lambda,\omega) = \dfrac{\alpha}{2\pi}K^{\text{soft}}(\lambda,\omega)\text{d}\hat{\sigma}^{\text{Born}}_{\lambda_{1}\lambda_{2}}.
\end{eqnarray*}
The infrared divergences of the soft photon contribution  compensates the corresponding divergences of the one-loop virtual QED RC.

$\bullet$ The hard photon contribution $\hat{\sigma}^{\mathrm{hard}}(\omega)$

We assume that the spinor formalism is more convenient for obtaining HAs of the hard photon Bremsstrahlung contribution $\hat{\sigma}^{\mathrm{hard}}(\omega)$. Below we describe in detail the process of obtaining the analytical gauge-independent form of HAs taking into account the masses $m_{i}$ of all particles for the
$4f \gamma \to 0$ process. 
Then we unfold HAs to the channel (\ref{partonllL}) for the massless case.

To calculate the hard photon Bremsstrahlung contribution we use the non-linear gauge~\cite{GAVELA1981257,Romao:1986vp}, since there is no interaction of the photon with charged would-be-Goldstones, and the vertex structures are more transparent.

Let $Q_i$ be the charge of a particle with momentum $p_i$. We also adopt notation $p_{i\dots j}=p_i+\dots+p_j$ and $Q_{i\dots j}=Q_i+\dots+Q_j$, $z_{i\dots j} = p_{i\dots j}^2 - (m_i +\dots+ m_j)^2$.
Obviously, $p_{12345}=0$ and $Q_{1234}=0$.
Using the Dirac algebra, we can allow one to separate the scalar part from the spin-induced part: 
$  (\gs p_{25} + m_2)\gs\varepsilon_5 u_2   = (2 \varepsilon_5\cdot p_2 + \gss F_5)u_2$, 
where $F_5^{\mu\nu}=p_5^{[\mu}\varepsilon_{5}^{\nu]}$ is the Maxwell bivector, and $\gss F_5 = \gamma_\mu\gamma_\nu F_5^{\mu\nu}$.

The amplitude is split into the sum of two gauge-independent contributions from initial (ISR) and final state radiation (FSR), 
\begin{align*} 
\mathcal{H}^{\text{hard}} &=  ie^2  \Big[ \dfrac{\chiW(\hat{s}_{34})}{\hat{s}_{34}} A^\mathrm{ISR}  + \dfrac{\chiW(\hat{s}_{12})}{\hat{s}_{12}}  A^\mathrm{FSR} \Big]
,\\ 
A^\mathrm{ISR}&= \dfrac{\Tr[\gs p_1\gs p_2\gss F_5]}{z_{15}+z_{25}}\bigg(
    \dfrac{Q_1}{z_{15}}  
-  \dfrac{ Q_2}{z_{25}}   
\bigg)  \calB_{12,43}  
\\
&+     Q_{2}\calB_{15,43}\calC_{52} - Q_{1}\calC_{15} \calB_{52,43} 
 - \dfrac{Q_{12}}{z_{15}+z_{25}}\calG_{5,12,43},
\\
A^\mathrm{FSR}&= \dfrac{\Tr[\gs p_3\gs p_4\gss F_5]}{z_{35}+z_{45}}\bigg(
    \dfrac{Q_3}{z_{35}}  
-  \dfrac{ Q_4}{z_{45}}  
\bigg)     \calB_{12,43}  
\\
&+   Q_{3}\calB_{12,45}\calC_{53} - Q_{4}\calC_{45}\calB_{12,53} 
-      \dfrac{Q_{34}}{z_{35}+z_{45}}\calG_{5,43,12}.
\end{align*}
The following building blocks are introduced:
\begin{align*}
\begin{aligned}
&\calB_{12,43} = \spVI{1} \gamma^\mu \omega_{-}\spIU{2} \otimes\spVI{4}\gamma_\mu \omega_{-} \spIU{3}
,\\
&\calG_{5,12,43} = 4 F_5^{\nu\mu}\;\spVI{1} \gamma_\mu \omega_{-}\spIU{2} \otimes \spVI{4}\gamma_\nu \omega_{-} \spIU{3}
,
\\
&\gss F_5 = \spIU{5} \spVI{5}
,
\;\calG_{5,12,43} = - \calG_{5,43,12}, 
\;\calC_{ij} = \dfrac{\spVI{i}\spIU{j}}{z_{ij}}.
\end{aligned}
\end{align*}
These blocks can easily be evaluated in the Weyl representation of $\gamma$-matrices.
Let us consider the massless case. Dirac spinors can be expressed in terms of Weyl ones:
\begin{align*}&
\spIU{2}^+ = \begin{pmatrix}
\SpIA{2}  \\
0  
\end{pmatrix}
,&&
\spIU{2}^- = \begin{pmatrix}
  0\\
 \SpIB{2}
\end{pmatrix}
,&&
\begin{aligned}
\spVI{1}^+ &= \begin{pmatrix}
\SpAI{1} & 0  
\end{pmatrix}
,\\
\spVI{1}^- &= \begin{pmatrix} 
0 & \SpBI{1}
\end{pmatrix}.
\end{aligned} 
\end{align*}
Applying the Fierz identities and taking into account the fact that in the massless case there are only two nonzero HAs,  we obtain
\begin{align*}
\calB^{+-+-}_{12,43} &
% = \SpAIIB{1}{\gsAB\sigma^\mu}{2}  \SpAIIB{4}{\gsAB\sigma^\mu}{3} 
=2\SpAIA{1}{4}\SpBIB{3}{2} 
,&&
\begin{aligned}
 \calG^{+,+-,+-}_{5,12,43} &= 
2  \SpAIA{5}{1}\SpBIB{2}{3}\SpAIA{4}{5}
,\\
 \calG^{-,+-,+-}_{5,12,43} &= 
2 \SpBIB{5}{2}\SpAIA{1}{4}\SpBIB{3}{5}.
\end{aligned}
\end{align*}
Substituting these expressions and performing further simplifications (using Schouten identity and momentum conservation), we finally obtain the following result:
\begin{align*}
% a^\mathrm{ISR}&= \Big(\dfrac{ Q_1}{z_{15}}  - \dfrac{ Q_2}{z_{25}} \Big) \dfrac{a^\mathrm{ISR}_2}{z_{15}+z_{25}}  
% \\
A^\mathrm{ISR}_{+,+-+-}&= 2\Big(\dfrac{ Q_1}{z_{15}}  - \dfrac{ Q_2}{z_{25}} \Big) \dfrac{ \SpAIA{5}{2} \SpAIA{5}{1} \SpAIA{3}{4} \SpBIB{2}{3}^2}{z_{15}+z_{25}},
\\
A^\mathrm{FSR}_{+,+-+-}&= 2\Big(\dfrac{ Q_3}{z_{35}}  - \dfrac{ Q_4}{z_{45}} \Big) \dfrac{ \SpAIA{5}{4} \SpAIA{5}{3} \SpAIA{1}{2} \SpBIB{2}{3}^2}{z_{35}+z_{45}},
\\
A^\mathrm{ISR}_{-,+-+-}&= 2\Big(\dfrac{ Q_1}{z_{15}}  - \dfrac{ Q_2}{z_{25}} \Big) \dfrac{ \SpBIB{1}{5}  \SpBIB{2}{5}  \SpBIB{3}{4} \SpAIA{4}{1}^2}{z_{15}+z_{25}},
\\
A^\mathrm{FSR}_{-,+-+-}&= 2\Big(\dfrac{ Q_3}{z_{35}}  - \dfrac{ Q_4}{z_{45}} \Big) \dfrac{ \SpBIB{3}{5} \SpBIB{4}{5} \SpBIB{1}{2} \SpAIA{4}{1}^2}{z_{35}+z_{45}}.
\end{align*}

\section{Spin-dependent observables \label{section2}}
 
We  evaluate several spin-dependent observables: 
polarized components of the total cross section $\sigma^{i}$, where
\begin{equation}
   \label{i} 
   i=\{00,\;++,\;--,\;+-,\;-+\};
\end{equation}
the parity-violating single-spin asymmetries ${A}_{\mathrm{L}}$, 
which are the basis of knowledge of polarized PDFs;
and the double-spin asymmetries ${A}_{\mathrm{LL}}$, which are the main test for the sea quark polarization. 
 
For convenience, we introduce the following combinations:
\begin{eqnarray*}
\sigma &=&
\frac{1}{4}\left(\sigma^{++}+\sigma^{+-}+\sigma^{-+}+\sigma^{--}\right),\\
\Delta\sigma_{\mathrm {L}} &=&
\frac{1}{4}\left(\sigma^{++}+\sigma^{+-}-\sigma^{-+}-\sigma^{--}\right),\\   
\Delta\sigma_{\mathrm {LL}} &=&
\frac{1}{4}\left(\sigma^{++}-\sigma^{+-}-\sigma^{-+}+\sigma^{--}\right),
\end{eqnarray*}
where $\sigma$ is unpolarized cross section.

Definitions of the single-spin asymmetry ${A}_{\mathrm{L}}(\eta_{\ell})$
and the double-spin asymmetry ${A}_{\mathrm{LL}}(\eta_{\ell})$
are
\begin{eqnarray}\label{AL}\label{ALL}
{\text{A}}_{\mathrm{L}(\mathrm{LL})}(\mathrm{\eta_{\ell}})
 = \dfrac{{ d(\Delta\sigma_{\mathrm {L}(\mathrm{LL})})}/{\ds d\mathrm{\eta_{\ell}}}}
    {{\ds d\sigma}/{\ds d\mathrm{\eta_{\ell}}}},
\end{eqnarray}

where pseudo-rapidity $\eta_{\ell}$ is 
\begin{eqnarray*}
\eta_{\ell}=-\ln\left(\tan \dfrac{\vartheta_{\ell}}{2}\right). 
\end{eqnarray*}

Here $\vartheta_{\ell}$ is the angle of the $\ell$
in the laboratory frame.
The $z$ axis is directed along the momentum of the first proton. 

The relative correction 
$\delta^{i}$ (in \%) is defined as
\begin{eqnarray}\label{deltai}
\delta^{i}=\dfrac{\sigma^{\mathrm{NLO}, i}}{\sigma^{\mathrm{LO}, i}}-1, \%,
\end{eqnarray}
where $i$ is components of cross section, see (\ref{i}).

\section{Comparison}

To test the validity of our calculation, we compared our results for the unpolarized case of CC DY with
independent NLO calculations realized in the programs {\tt HORACE} \cite{CarloniCalame:2003ux,CarloniCalame:2006zq}, {\tt WGRAD2} \cite{Baur:1998kt} and { \tt SANC} \cite{Arbuzov:2005dd}. 
The results of a tuned comparison of the kinematic distributions at the NLO level, as shown in section 4.4 \cite{TeV4LHC-Top:2007fwh}, demonstrate that the events modeled using the {\tt SANC} approach correctly reproduce the NLO distributions of the other two codes.

In this paper, we use the spinor formalism to obtain the HAs of the hard photon bremsstrahlung contribution, in contrast to the standard approach of matrix element squaring used in previous calculations of the hard cross section $\sigma^{\text{hard}}$ \cite{Bardin:2012jk,Bondarenko:2013nu,TeV4LHC-Top:2007fwh}. The computational module for $\sigma^{\text{hard}}$ based on the spinor formalism has been thoroughly tested, and its results are in agreement with those obtained using the standard approach.

The one-loop correction for polarized CC Drell-Yan processes was first estimated in the papers \cite{Zykunov:2001mn,Zykunov:2003gm} using a semi-analytical approach. In this study, we achieved excellent agreement for unpolarized distributions at the Born level with the input parameter set used here. However, single- and double-spin asymmetries don't match.

\section{Numerical results \label{section3}}

We investigate the effect of one-loop EW RCs on the spin-dependent observables
in the $G_{\mu}$ scheme at the energy $\sqrt{s} = 510$ GeV in the input parameter set adopted in {\tt SANC}. The input parameters set is given below.

\begin{table}[!h]
    \centering
    \begin{tabular}{lcllclc}
    $\alpha^{-1}(0)$ &=& 137.035999084,    & & && \\
     $G_F$           &=& 1.1663787$\times10^{-5}$ GeV$^{-2}$, & & & &\\
    $M_W$            &=& 80.379 GeV,       & $\Gamma_W$ &=& 2.085 GeV,                           & \\
    $M_Z$            &=& 91.1876 GeV,      & $\Gamma_Z$ &=& 2.4952 GeV,                          & \\
    $M_H$            &=& 125.25 GeV,       &                       & \\
    $m_e$            &=& 0.51099895 MeV,   &                        & \\
    $m_\mu$          &=& 0.1056583745 GeV, & $m_\tau$   &=& 1.77686 GeV,                         & \\
    $m_u$            &=& 0.066 GeV,        & $m_d$      &=& 0.066 GeV,                           & \\
    $m_c$            &=& 1.67 GeV,         & $m_s$      &=& 0.15 GeV,                            & \\
    $m_t$            &=& 172.76 GeV,       & $m_b$      &=& 4.78 GeV,                            & \\
    $|V_{ud}|$       &=& 0.9737,           & $|V_{cd}|$ &=& 0.221,                               & \\
    $|V_{us}|$       &=& 0.2252,           & $|V_{cs}|$ &=& 0.987,                               & \\
    $|V_{ub}|$       &=& 0,                & $|V_{cb}|$ &=& 0.                                   & \\
    \end{tabular}
    \label{parameters_set}
\end{table}

The following cuts were applied ($\ell = e, \; \mu$):
\begin{eqnarray*}
\label{Cuts}
&&W^{+}: p_\perp(\ell^+) > 25 \; \text{GeV}, \; p_\perp(\nu_\ell) > 25 \; \text{GeV}, \nonumber 
  \;  |\eta(\ell^+)| < 2.5, \; M(\ell^+\nu_\ell) > 1 \; \text{GeV}, \nonumber \\
&&W^{-}: p_\perp(\ell^-) > 25 \; \text{GeV}, \; p_\perp(\bar{\nu}_\ell) > 25 \; \text{GeV}, \nonumber 
   \; |\eta(\ell^-)| < 2.5, \; M(\ell^-\bar{\nu}_\ell) > 1 \; \text{GeV}.
\end{eqnarray*}
 
We used the PDF set {\tt NNPDF23\_nlo\_as\_0119} 
for the unpolarized parton distributions $f_{q_i}$ and the PDF set {\tt NNPDFpol11\_100} for the longitudinally polarized ones $\Delta f_{q_i}$ 
from the {\tt LHAPDF6} library with the factorization scale 
$\mu_F = M_{\ell\ell}$~\cite{ParticleDataGroup:2020ssz}.

We present the distributions 
for the polarized components of the Born cross section (LO)
and corresponding $\delta^{i},\;\%$ (\ref{deltai}), % the parity violating 
single-spin $\text{A}_\mathrm{L}$ (\ref{AL}), double-spin $\text{A}_\mathrm{LL}$ (\ref{ALL}) asymmetries and the corresponding $\Delta \text{A}_\text{L} = \text{A}_\text{L}^{\text{NLO}}-\text{A}_\text{L}^{\text{LO}}$ over pseudo-rapidity at one-loop EW level (NLO EW).

\begin{figure}
    \begin{minipage}{1\linewidth}
    \begin{center}
        \includegraphics[width=0.65\linewidth]{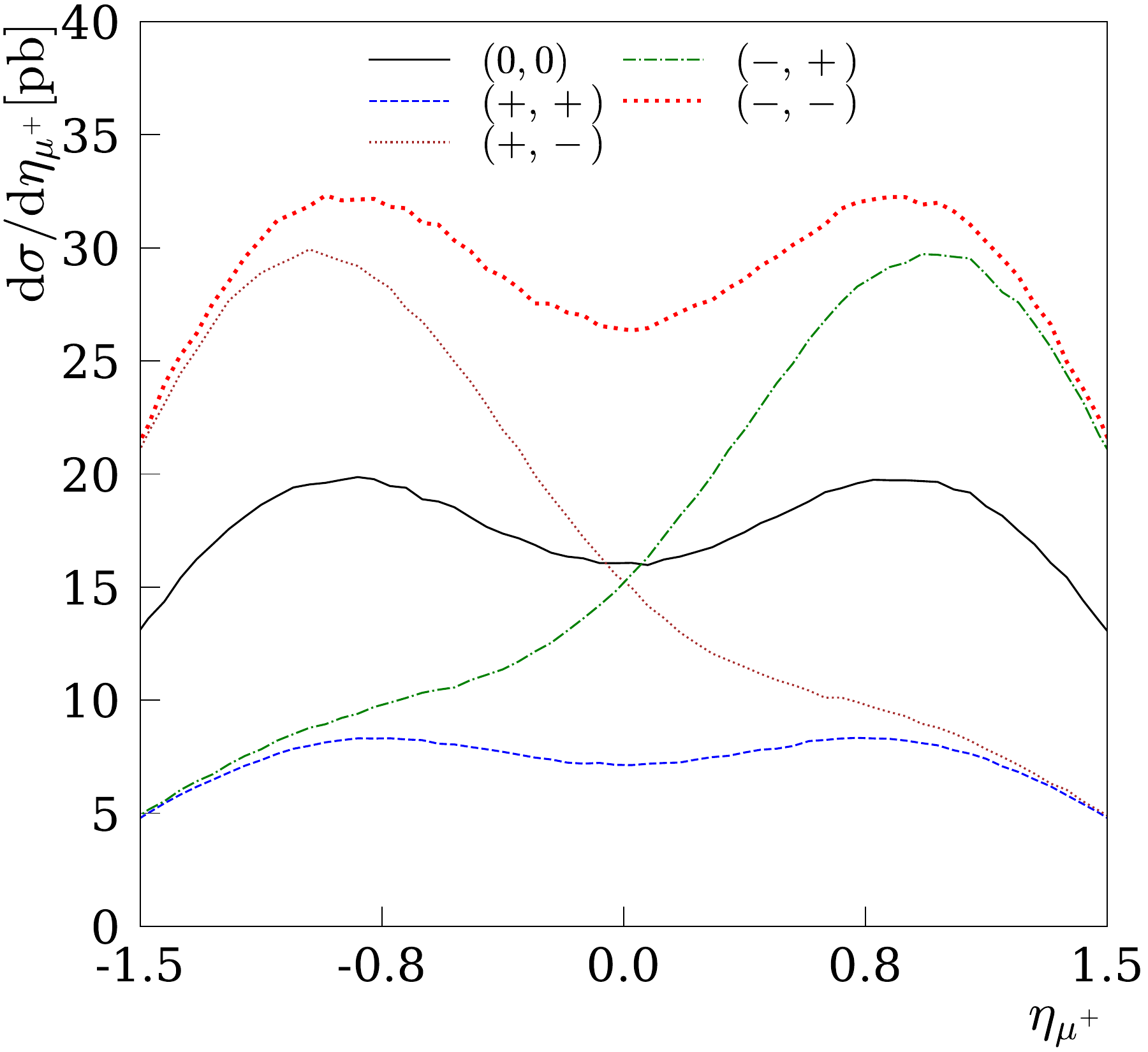} 
        \caption{The distribution for the Born level cross section in pb 
        over pseudo-rapidity $\eta_{\mu^{+}}$.} 
        \label{fig:cs_CC_antimu}
    \end{center}
    \end{minipage}
    \begin{minipage}{1\linewidth}
    \begin{center}
        \includegraphics[width=0.65\linewidth]{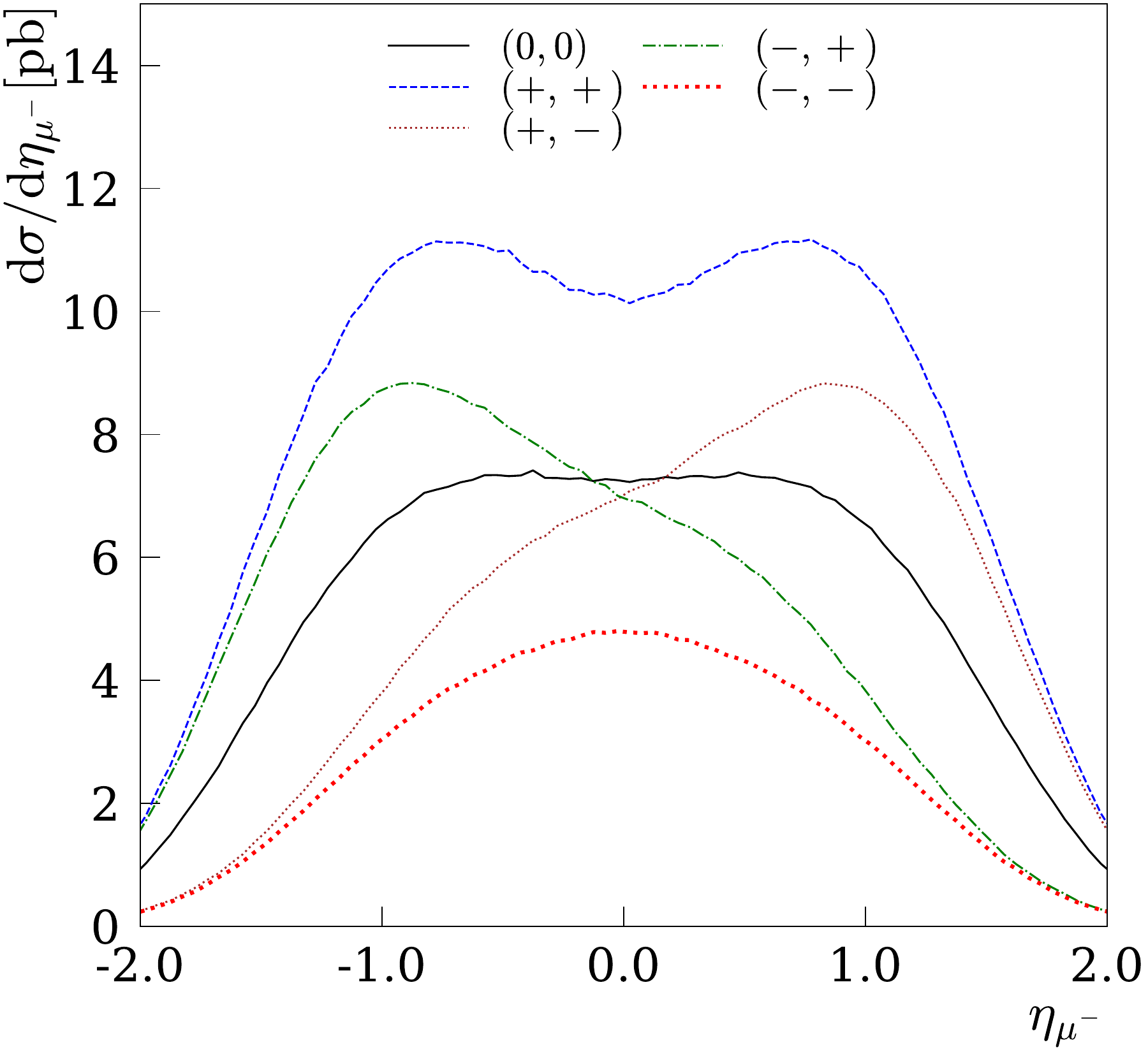}
        \caption{The same as in FIG. \ref{fig:cs_CC_antimu} but over the  pseudo-rapidity $\eta_{\mu^{-}}$. }
        \label{fig:cs_CC_mu}  
    \end{center}
    \end{minipage}
\end{figure}

For the distributions of polarized components of the Born cross section we demonstrate significant shifts 
with respect to unpolarized case the $W^{+}$ channel (FIG. \ref{fig:cs_CC_antimu}) and the $W^{-}$ channel (FIG. \ref{fig:cs_CC_mu}).

\begin{figure}
    \begin{minipage}{1\linewidth}
    \begin{center}
        \includegraphics[width=0.65\linewidth]{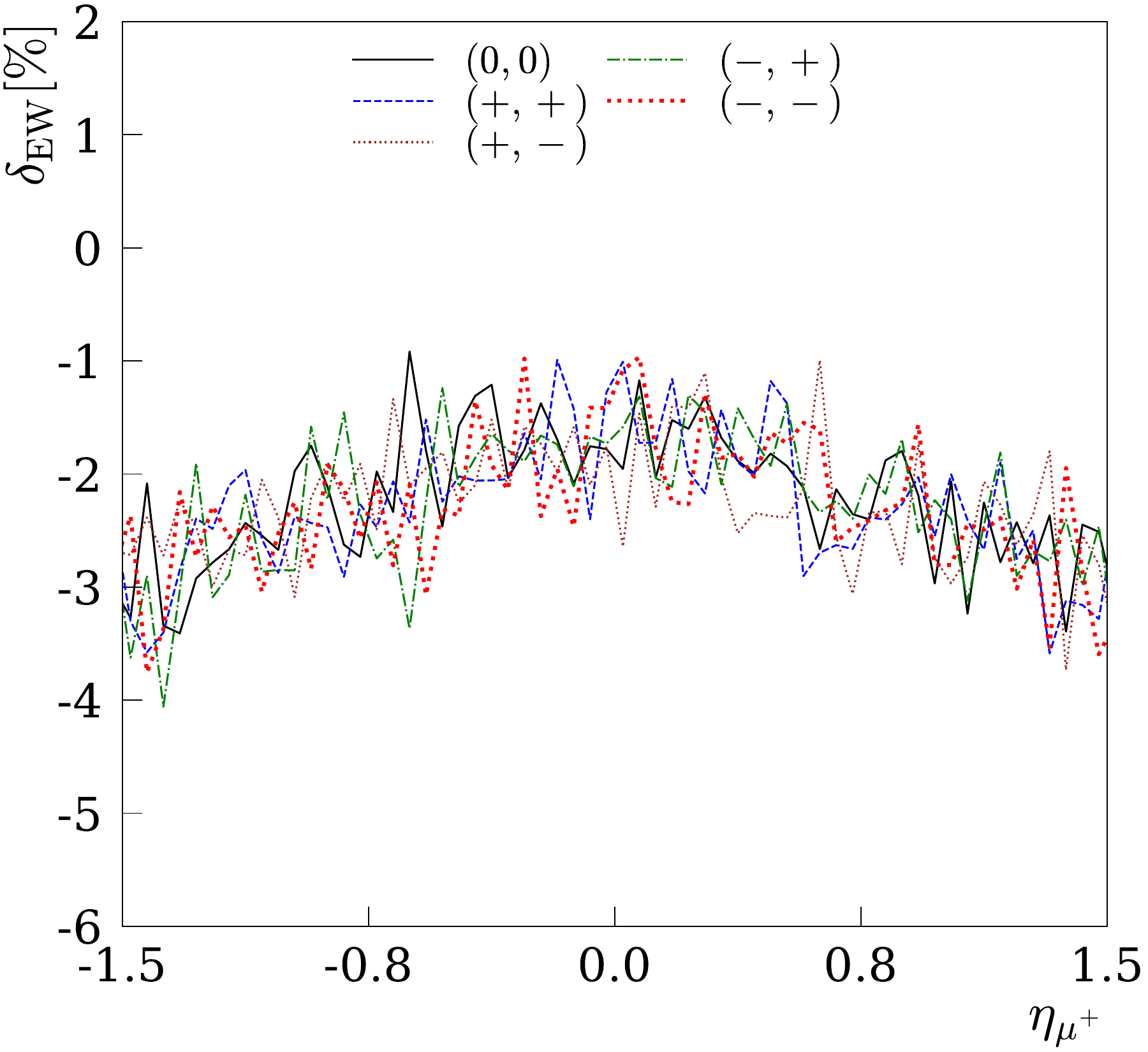}
        \caption{The relative corrections $\delta^{i}$ in \% 
        over pseudo-rapidity $\eta_{\mu^{+}}$.}
        \label{fig:delta_CC_antimu}
    \end{center}
    \end{minipage}
    \begin{minipage}{1\linewidth}
    \begin{center}
        \includegraphics[width=0.65\linewidth]{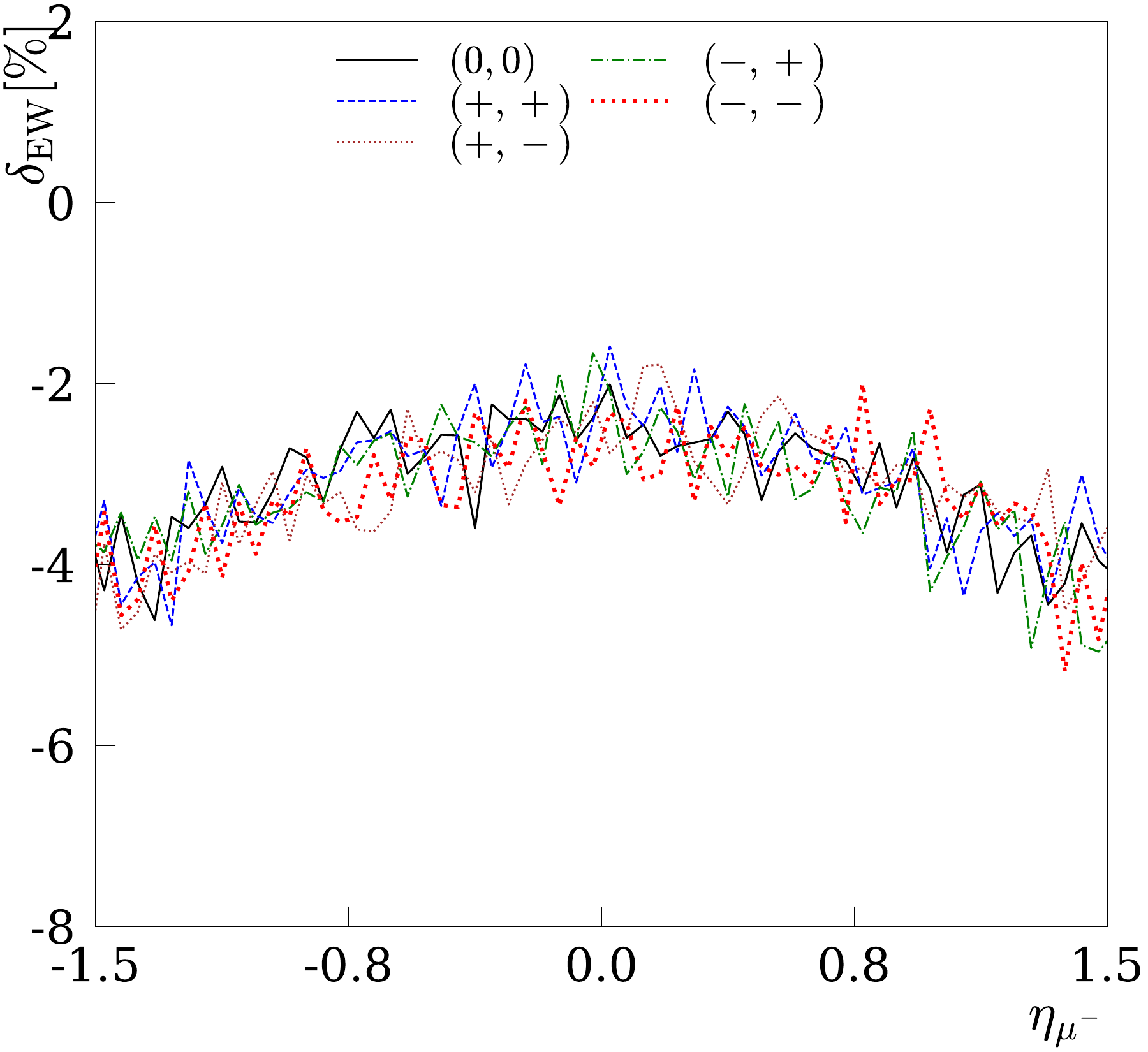}
        \caption{The same as in FIG. \ref{fig:delta_CC_antimu} but over the  pseudo-rapidity $\eta_{\mu^{-}}$.}
        \label{fig:delta_CC_mu}
    \end{center}
    \end{minipage}
\end{figure}

Corresponding relative corrections $\delta^{i}$ (\ref{deltai}) are shown in the FIG. \ref{fig:delta_CC_antimu} for the $W^{+}$ channel and in the FIG. \ref{fig:delta_CC_mu} for the $W^{-}$ channel. 
The ranges of relative corrections $\delta^{i}$ are approximately from $-4\%$ to $-1\%$ and from $-4\%$ to $-2\%$, for the corresponding channels.
We see that the impact of the EW RCs is small.

\begin{figure}
    \begin{minipage}{1\linewidth}
    \begin{center}
        \includegraphics[width=0.6\linewidth]{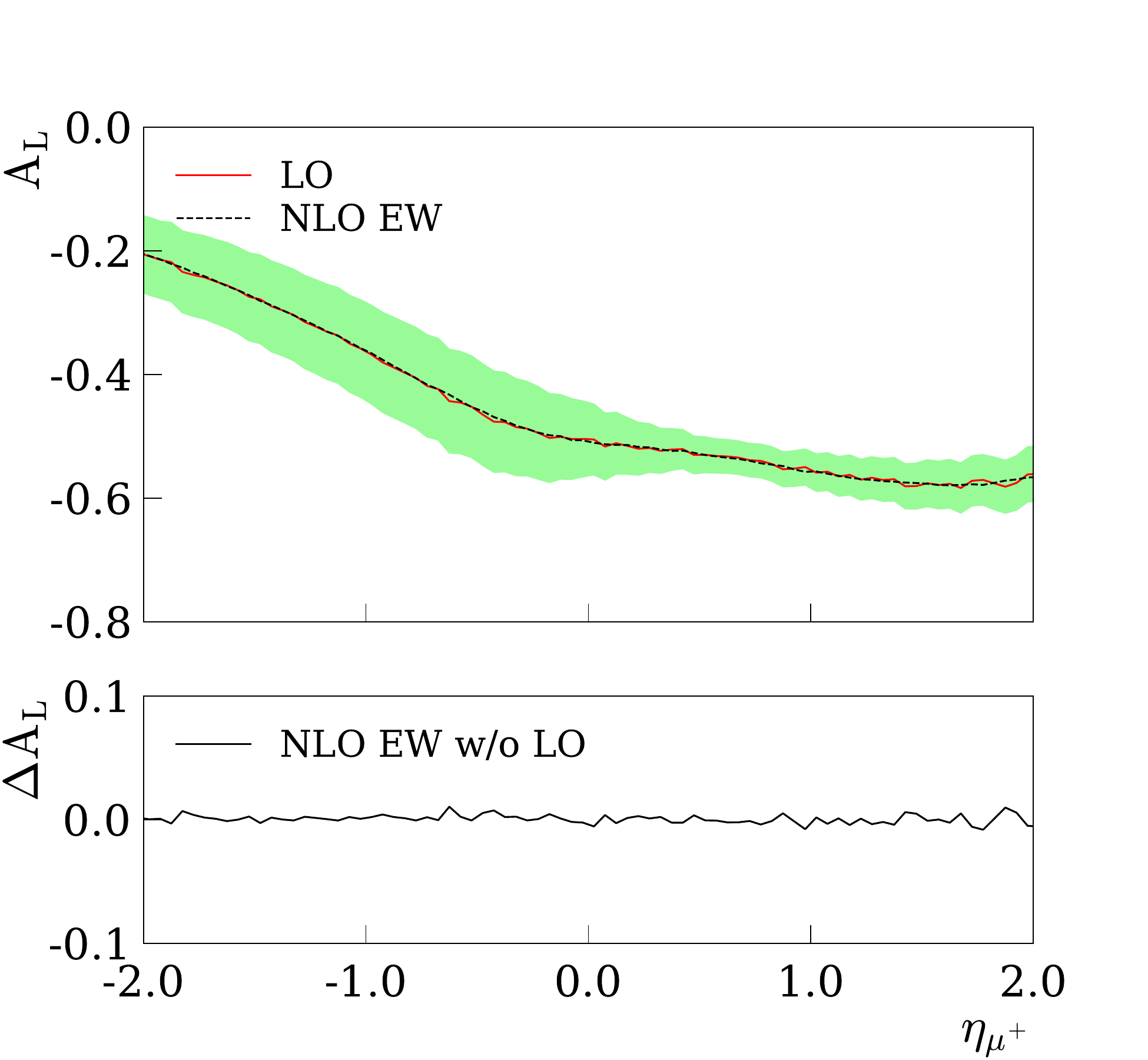} 
        \caption{The single-spin asymmetry $\text{A}_{\text{L}}$ 
        with the green 1$\sigma$ PDF uncertainty band at the LO 
        and NLO EW levels (upper panel)
        and corresponding difference $\Delta \text{A}_{\text{L}}$ (bottom panel) over the pseudo-rapidity $\eta_{\mu^{+}}$.}
        \label{fig:AL_CC_antimu}
    \end{center}
    \end{minipage}
    \begin{minipage}{1\linewidth}
    \begin{center}
        \includegraphics[width=0.6\linewidth]{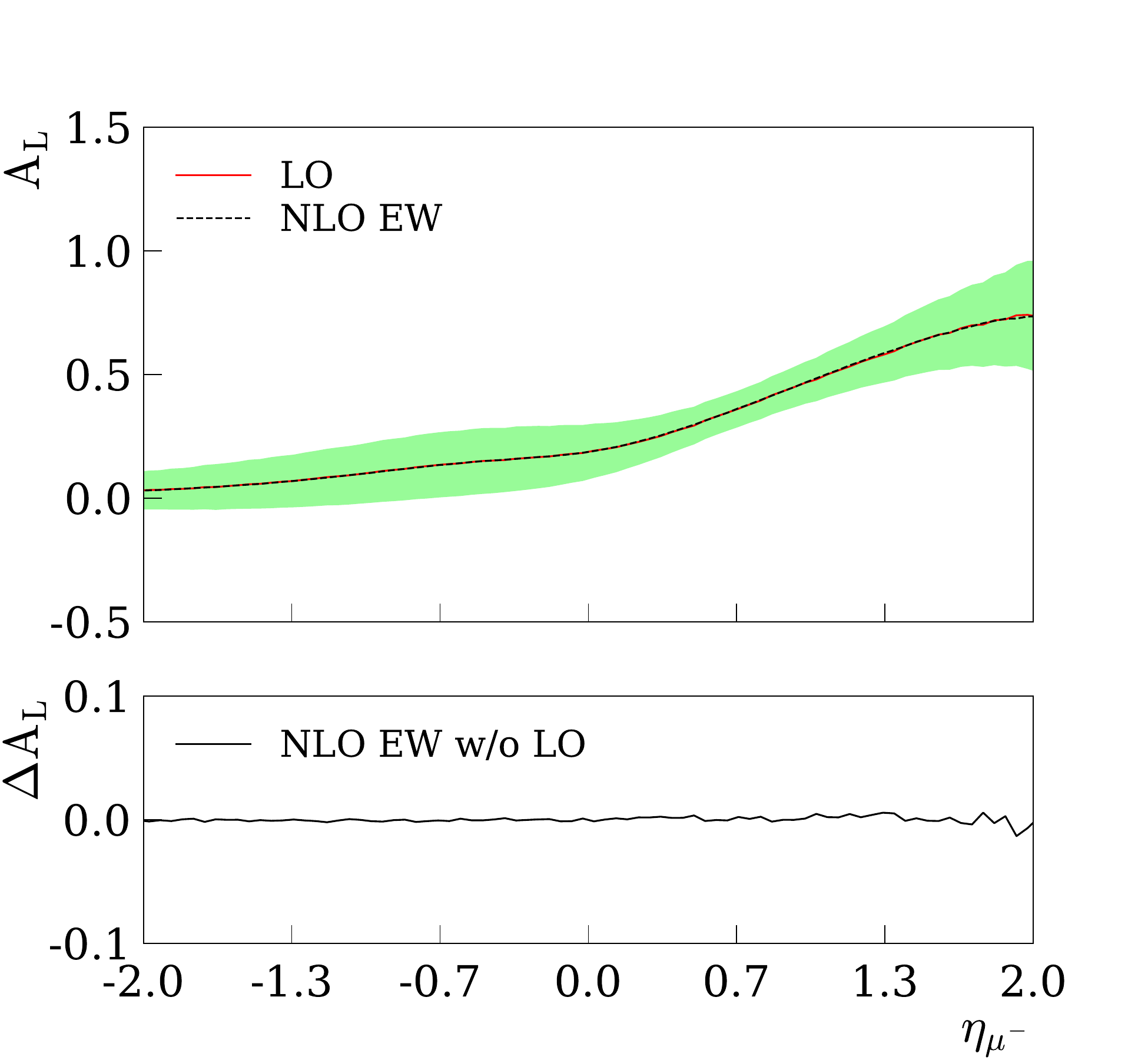}
        \caption{The same as in FIG. \ref{fig:AL_CC_antimu} but over the  pseudo-rapidity $\eta_{\mu^{-}}$.}
        \label{fig:AL_CC_mu}
    \end{center}
    \end{minipage}
\end{figure}

The contribution of the EW RCs doesn't have a significant impact on the single-spin asymmetry for the channel $W^{+}$ (FIG. \ref{fig:AL_CC_antimu}) and for the channel $W^-$ (FIG. \ref{fig:AL_CC_mu}).
Additionally, an uncertainty band is calculated using one hundred replicas of the PDF set {\tt NNPDFpol11\_100}.
The red line corresponds to the central value and the green band covers the $1\sigma$ PDF uncertainty.
% In some region the NLO line goes beyond the band, i.e. the EW RCs are important in this region.  

\begin{figure}
    \begin{minipage}{1\linewidth} 
    \begin{center}
    \includegraphics[width=0.7\linewidth]{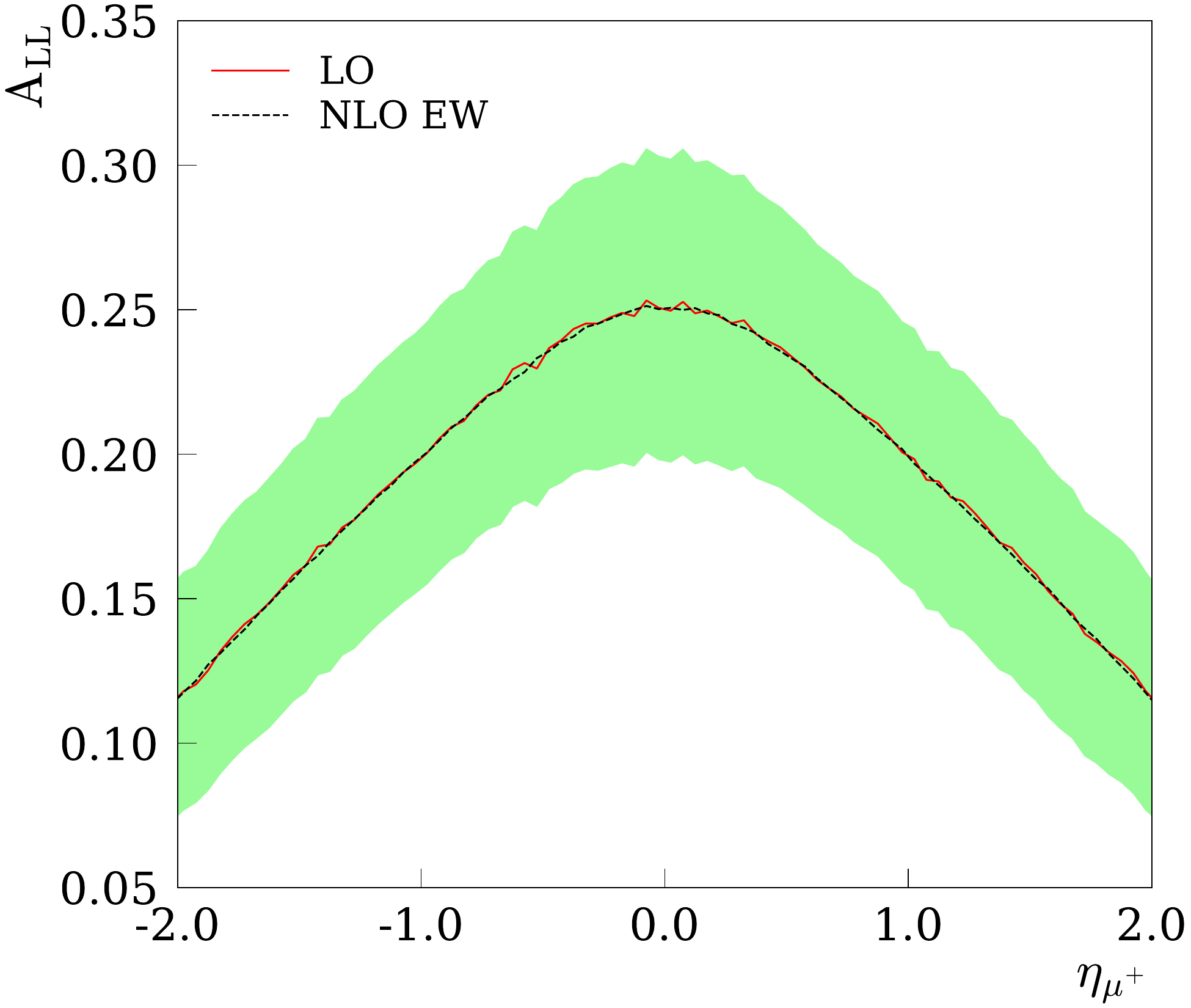}
        \caption{The double-spin asymmetry $\text{A}_{\text{LL}}$ with the green 1$\sigma$ PDF uncertainty band at the LO and NLO EW levels over the pseudo-rapidity $\eta_{\mu^{+}}$.}
        \label{fig:ALL_CC_antimu}
    \end{center}
    \end{minipage}    
    \begin{minipage}{1\linewidth}
    \begin{center}
         \includegraphics[width=0.7\linewidth]{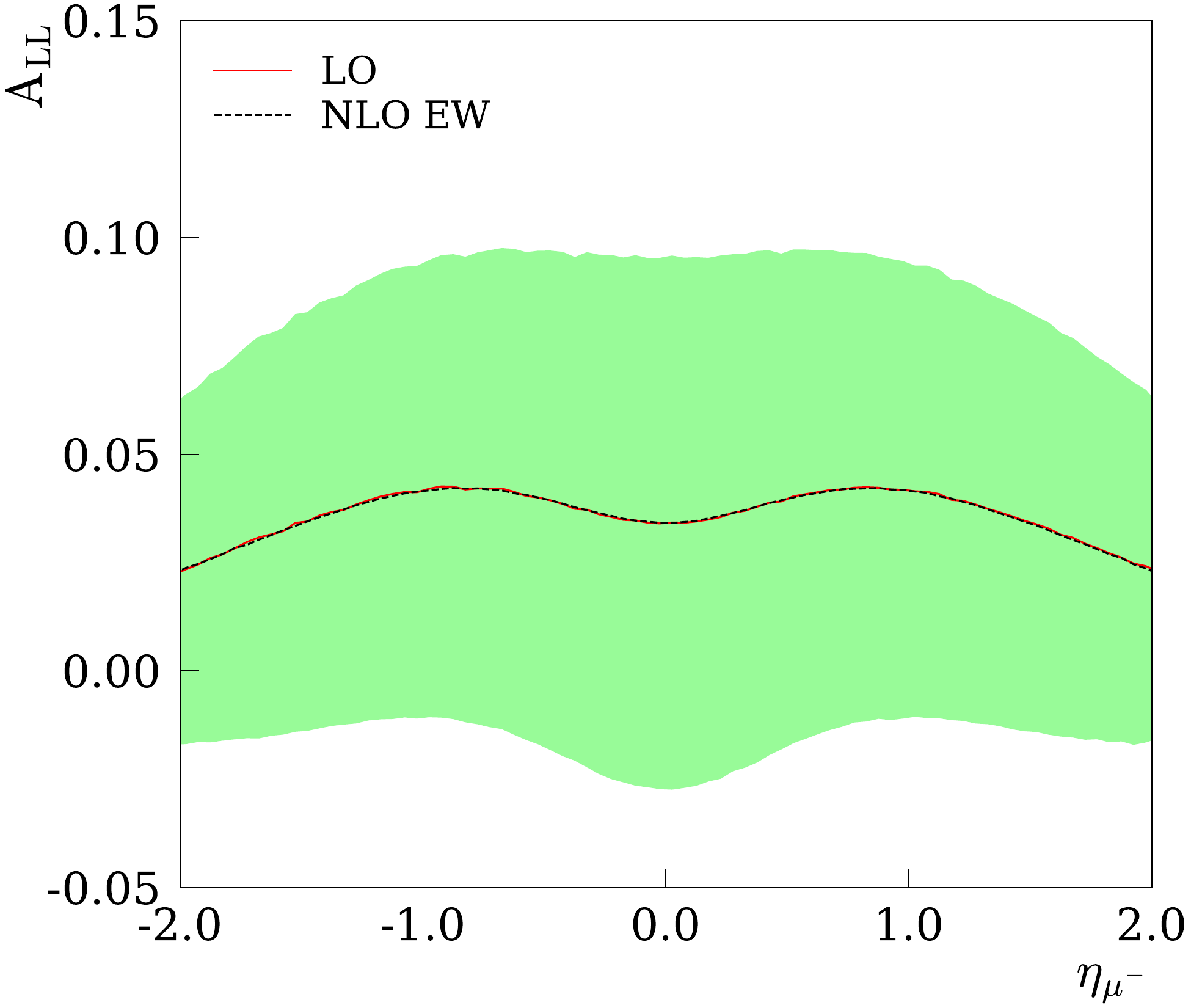}
        \caption{The same as in FIG. \ref{fig:ALL_CC_antimu} but over the  pseudo-rapidity $\eta_{\mu^{-}}$.}
        \label{fig:ALL_CC_mu}
    \end{center}
    \end{minipage}
\end{figure}

For double-spin asymmetry $\text{A}_{\text{LL}}$ (FIG. 
\ref{fig:ALL_CC_antimu}-\ref{fig:ALL_CC_mu}) 
the NLO contribution is also negligible but the uncertainty band associated with the impact of PDF uncertainties on the observable is very wide.

\section{Conclusions \label{Conclusion}} 

We studied the sensitivity of the observables 
for $W^\pm$ boson production in longitudinally polarized $pp$-collisions to the one-loop EW RCs within the RHIC experiment conditions.
The impact of the one-loop EW RCs
to the cross sections is small and to single- and double-spin asymmetries is negligible with certain experimental cuts. 

\section{Funding}
\label{sec:funding}
The research is supported by the Russian Science Foundation, project No. 22-12-00021.


\clearpage

\end{document}